\documentclass{ifacconf}

\usepackage{graphicx}      
\usepackage{natbib}        

\usepackage{pgf}
\usepackage{tikz}
\usepackage{xcolor}
\usepackage{amsmath} 
\usepackage{amssymb}  
\usepackage{algorithm}
\usepackage{algorithmic}
\usepackage{textcomp}
\usepackage{amsfonts}
\usepackage{graphicx}
\usepackage{epstopdf}
\usepackage{mathrsfs}
\usepackage{color}

\providecommand{\norm}[1]{\lVert#1\rVert}
\newcommand{\IM}{\mathrm{im}}

\newcommand{\SPAN}{\mathrm{span}}
\newtheorem{Definition}{Definition}
\newtheorem{Theorem}{Theorem}
\newtheorem{Lemma}{Lemma}

\newtheorem{Corollary}{Corollary}
\newtheorem{Remark}{Remark}
\newtheorem{Notation}{Notation}

\begin{document}
\begin{frontmatter}

\title{Model Reduction for Aperiodically Sampled Data Systems\thanksref{footnoteinfo}} 

\thanks[footnoteinfo]{This work was partially supported by ESTIREZ project of Region Nord-Pas de Calais, France and by ANR project ROCC-SYS (ANR-14-CE27-0008).}

\author[First]{Mert Ba\c{s}tu\u{g}} 
\author[First]{Laurentiu Hetel} 
\author[First]{Mih\'{a}ly Petreczky}

\address[First]{Centre de Recherche en Informatique, Signal et Automatique de Lille (CRIStAL), UMR CNRS 9189, Ecole Centrale de Lille, 59650 Villeneuve d'Ascq, France (e-mail: mert.bastug@univ-lille1.fr, laurentiu.hetel@ec-lille.fr, mihaly.petreczky@ec-lille.fr).}

\begin{abstract}                
Two approaches to moment matching based model reduction of aperiodically sampled data systems are given. The term ``aperiodic sampling'' is used in the paper to indicate that the time between two consecutive sampling instants can take its value from a pre-specified finite set of allowed sampling intervals. Such systems can be represented by discrete-time linear switched (LS) state space (SS) models. One of the approaches investigated in the paper is to apply model reduction by moment matching on the linear time-invariant (LTI) plant model, then compare the responses of the LS SS models acquired from the original and reduced order LTI plants. The second approach is to apply a moment matching based model reduction method on the LS SS model acquired from the original LTI plant; and then compare the responses of the original and reduced LS SS models. It is proven that for both methods, as long as the original LTI plant is stable, the resulting reduced order LS SS model of the sampled data system is quadratically stable. The results from two approaches are compared with numerical examples.
\end{abstract}


\end{frontmatter}

\section{Introduction}

The topic of model reduction deals with computing simpler approximation models for an original complex model [\cite{antoulas}]. For system classes which can be represented by state-space (SS) models, the ``complexity'' of a model refers usually to the state-space dimension of the corresponding model. Hence a ``simpler model'' is a model with less number of states whose input-output behavior is close to the one of the original system. In this paper, the term model reduction is used in this sense, i.e., approximating the input-output behavior of an original SS model with an another SS model with less number of states. 

Aperiodically sampled data systems appear commonly in applications since they can be used for modeling various phenomena encountered in the context of large scale networked and embedded control systems [\cite{Hespanha_TNCS_2007,Hristu_Varsakelis_Birkhauser_2003,Zhang_CSM_2001,hetel_auto_survey, brockett_1997, tabuada_heemels}]. In turn, the dimension of corresponding SS models for such systems can be very big due to the interaction of different subsystems in the network. Simulations for control synthesis or performance specifications regarding the output behavior can easily become intractable due to the complexity of the original model. Hence, model reduction approaches for such systems can be of great importance.

The paper states two model reduction procedures based on moment matching for aperiodically sampled data systems. Model reduction for sampled data systems has been considered previously on [\cite{barb_weiss,shieh_chang}]. Both papers deal with the case of periodical sampling and are valid only for the case when the considered plant is stable. In contrast, in the present paper the general aperiodic sampling case considered and the considered plant is allowed to be unstable. 

In this paper, the sampling interval is considered to be time varying and assumed to be taking its values from a finite set of possible sampling intervals. The input-output behavior at sampling instants of such sampled data systems can be modeled by discrete-time linear switched (LS) SS representations [\cite{gu_kharitonov_chen, zhang_2001, hetel_heemels_book_chapter}]. One of the model reduction procedures considered in the paper can be summarized as follows: Apply a classical moment matching algorithm to the original continuous-time linear time-invariant (LTI) plant to get a reduced order model, and then get an LS SS representation to model the aperiodically sampled system. The other approach given is to apply an analogous moment matching based model reduction algorithm to the LS SS representation which is computed from the original LTI plant. Since the sampling interval at each time instant acts as an additional control input for aperiodically sampled data systems, intuitively, a method of approximating the input-output behavior of such systems should make use of the information of the allowed sampling interval set. The second approach is given in accordance with this idea. For both of the approaches, it is shown that the resulting reduced order discrete-time LS SS representation of the sampled data system will be quadratically stable as long as the original continuous-time LTI model is stable.

The paper is organized as follows: In Section \ref{sect:sampled_to_LS}, we present the procedure of modeling an aperiodically sampled data system with an LS SS representation. In Section \ref{sect:first_approach} we present a brief overview of the concept of model reduction by moment matching for LTI SS representations and present the first model reduction approach in detail. In Section \ref{sect:second_approach} we briefly review the concept of model reduction by moment matching for LS SS representations and present the second model reduction approach in detail. In Section \ref{sect:conserv_stability} we show that the proposed methods preserve stability. In Section \ref{sect:numerical_examples} we illustrate the two approaches and compare their performances with two numerical examples.

\begin{Notation}
	In the following, we will use $\mathbb{Z}$, $\mathbb{N}$ and $\mathbb{R}_+$ to denote respectively the set of integers; the set of natural numbers including $0$ and the set $[0,+\infty)$ of nonnegative real numbers. We will use $I_a$ to denote the $a \times a$ identity matrix with $a \in \mathbb{N} \backslash \{ 0 \} $.
\end{Notation}

\section{Modeling of Sampled Data Systems with LS SS Representations} \label{sect:sampled_to_LS}

In this section we present briefly the process of modeling an aperiodically sampled continuous-time LTI system with a discrete-time LS SS model. We start with the formal definition of continuous-time LTI state-space (SS) representations.

An LTI SS representation $\Sigma_{\mathrm{LTI}}$ is a tuple $\Sigma_{\mathrm{LTI}}=(A,B,C)$ with $A \in \mathbb{R}^{n \times n}$, $B \in \mathbb{R}^{n \times m}$, $C \in \mathbb{R}^{p \times n}$. The state $x(t) \in \mathbb{R}^n$ and the output $y(t) \in \mathbb{R}^p$ of the LTI system $\Sigma_{\mathrm{LTI}}$ at time $t \geq 0$ is defined by \footnote{Unless stated otherwise, we take $x(0)=x_0=0$ for all classes of systems discussed in the paper for notational simplicity. Note that the result of the paper can easily be extended to the case of non-zero initial states.}
\begin{equation} \label{eq:LTI_cont_sys}
\Sigma_{\mathrm{LTI}}
\begin{cases} 
& \dot{x}(t) = Ax(t) + Bu(t) \\
& y(t) = Cx(t), \mbox{ } \forall t \in \mathbb{R}_+. 
\end{cases}
\end{equation}
In the following, $\dim(\Sigma_{\mathrm{LTI}})$ will be used to denote the dimension $n$ of the state-space of $\Sigma_{\mathrm{LTI}}$ and the number $n$ will be called the \emph{order} of $\Sigma_{\mathrm{LTI}}$.

Let $\Sigma_{\mathrm{LTI}}=(A,B,C)$ be a continuous-time LTI SS representation of the form \eqref{eq:LTI_cont_sys}. Let the state $x(t_k)$ and output $y(t_k)$ of $\Sigma_{\mathrm{LTI}}$ be sampled in arbitrary time instants $t_k$, $k \in \mathbb{N}$ such that $t_0=0$ and $t_{k+1}-t_k \in \mathcal{H}=\{ \hat{h}_1, \dots, \hat{h}_D \}$, $\hat{h}_1, \dots, \hat{h}_D \in \mathbb{R}_+$ for all $k \in \mathbb{N}$ to form the constant control signal $u(t)=u_k$ for all $t \in [t_k,t_k+1)$, $k \in \mathbb{N}$. Note that the sequence $t_k$, $k \in \mathbb{N}$ is monotonically increasing. The resulting sampled data system $\Sigma_{\mathrm{SD}}$ can be represented as follows:
\begin{equation} \label{eq:sampled_plant}
\Sigma_{\mathrm{SD}}
\begin{cases} 
& \dot{x}(t)= Ax(t) + Bu_k, \mbox{ } t \in [t_k, t_{k+1}), \mbox{ } k \in \mathbb{N} \\
& y_k = Cx(t_k) \\
& t_{k+1}=t_k + h_k, \mbox{ } h_k \in \mathcal{H}=\{ \hat{h}_1, \dots, \hat{h}_D \}.
\end{cases}
\end{equation}
In \eqref{eq:sampled_plant}, $x(t) \in \mathbb{R}^n$ is the state, $u_k \in \mathbb{R}^m$ is the constant input and $y(t) \in \mathbb{R}^p$ is the output at time $t \in \mathbb{R}_+$; $A \in \mathbb{R}^{n \times n}$, $B \in \mathbb{R}^{n \times m}$ and $C \in \mathbb{R}^{p \times n}$ are the same as the system parameters $(A,B,C)$ of $\Sigma_{\mathrm{LTI}}$. We call the set $\mathcal{H}=\{ \hat{h}_1, \dots, \hat{h}_D \}$, as the finite \emph{sampling interval set} and the value $h_k \in \mathcal{H}$ as the \emph{$k$th sampling interval}. We will use the shorthand notation $\Sigma_{\mathrm{SD}}=(A,B,C,\mathcal{H})$ for the sampled data system of the form \eqref{eq:sampled_plant}. Note that different from the model \eqref{eq:LTI_cont_sys}, model \eqref{eq:sampled_plant} has also $h_k \in \mathcal{H}$, $k \in \mathbb{N}$ as the control parameter in addition to the input $u(t)=u_k, \mbox{ } t \in [t_k, t_{k+1}), \mbox{ } k \in \mathbb{N}$.

The state $x_k=x(t_k)$ and output $y_k=y(t_k)$ of the sampled data system $\Sigma_{\mathrm{SD}}$ in \eqref{eq:sampled_plant} at sampling instants $t_k$, $k \in \mathbb{N}$ can be written by induction as
\begin{equation} \label{eq:cont_dyn_2}
\begin{aligned}
x_{k+1} & =x(t_{k+1}) = e^{Ah_k}x_k+ \left( \int_{0}^{h_k} e^{As} ds \right) Bu_k, \mbox{ } \forall k \in \mathbb{N}, \\
y_k & = Cx_k.
\end{aligned}
\end{equation}
Let 
\begin{equation} \label{eq:theta}
\Theta(h_k)=\int_{0}^{h_k} e^{As} ds.
\end{equation}
It is easy to see that the following holds:
\begin{equation} \label{eq:relation_exp_int}
e^{Ah_k}=I_n+A \Theta(h_k).	
\end{equation}
Replacing \eqref{eq:relation_exp_int} in \eqref{eq:cont_dyn_2} and defining the matrix functions $\Phi : \mathcal{H} \rightarrow \mathbb{R}^{n \times n}$ and $\Gamma : \mathcal{H} \rightarrow \mathbb{R}^{n \times m}$ as $\Phi(h_k)=e^{Ah_k}=I_n+A \Theta(h_k)$ and $\Gamma(h_k)=\Theta(h_k)B$, \eqref{eq:cont_dyn_2} can be rewritten as
\begin{equation} \label{eq:time_varying_repr}
\Sigma_{\mathrm{disc}} \left\{
\begin{aligned}
x_{k+1} & =\Phi (h_k)x_k + \Gamma (h_k) u_k,   \\
y_k & = Cx_k, \mbox{ } \forall h_k \in \mathcal{H}, \mbox{ } k \in \mathbb{N}.
\end{aligned} \right.
\end{equation}
With equation \eqref{eq:time_varying_repr}, the sampled LTI plant $\Sigma_{\mathrm{SD}}$ in \eqref{eq:sampled_plant} is modeled by a discrete-time, time-varying linear system $\Sigma_{\mathrm{disc}}$ whose read-out map (map represented by the matrix $C$) is time-invariant. Here, the discrete time instants $k \in \mathbb{N}$ of \eqref{eq:time_varying_repr} corresponds to the time instants $t_k \in \mathbb{R}_+$, $k \in \mathbb{N}$ for the original sampled data system $\Sigma_{\mathrm{SD}}$. In addition, the state $x_k$ and the output $y_k$ of \eqref{eq:time_varying_repr} corresponds to the state $x(t_k)$ and output $y(t_k)$ of $\Sigma_{\mathrm{SD}}$ at the sampling instants $t_k  \in \mathbb{R}_+$ when $u(t)=u_k$ for $t \in [t_k,t_{k+1})$. Hence we have built the relationship between the sampled data system $\Sigma_{\mathrm{SD}}=(A,B,C,\mathcal{H})$ and the corresponding discrete-time, linear time-varying system representation $\Sigma_{\mathrm{disc}}$.

Since the sampling interval $h_k$ between any two consecutive sampling instants can take its values only from the finite set $\mathcal{H}=\{ \hat{h}_1, \dots, \hat{h}_D \}$ one approach to design control for the model \eqref{eq:time_varying_repr} is to create an LS SS model from \eqref{eq:time_varying_repr}. The idea is that since the set $\mathcal{H}$ has $D$ elements, $\Theta(h_k)$ can only take $D$ different values for all $k \in \mathbb{N}$. In turn, \eqref{eq:time_varying_repr} can be used to create an LS SS representation with $D$ discrete modes. Below we summarize this procedure.

\begin{Notation}
	Let $a,b \in \mathbb{N}$. In the following, we use $\mathbb{I}_{a}^{b}$ to denote the set $\mathbb{I}_{a}^{b}=\{ c \in \mathbb{N} \mid a \leq c \leq b \}$.
\end{Notation}

Let the matrices $\hat{A}_1, \dots, \hat{A}_D \in \mathbb{R}^{n \times n}$ and $\hat{B}_1, \dots, \hat{B}_D \in \mathbb{R}^{n \times m}$ be defined by

\begin{equation} \label{eq:LSS_parameters}
\begin{aligned}
	\hat{A}_i & =I_n + A \Theta(\hat{h}_i), \mbox{ } \forall i \in \mathbb{I}_1^{D}, \\
	\hat{B}_i & =\Theta(\hat{h}_i)B, \mbox{ } \forall i \in \mathbb{I}_1^{D}.
\end{aligned}
\end{equation}

Using \eqref{eq:LSS_parameters}, \eqref{eq:time_varying_repr} can be rewritten as the following SS representation

\begin{equation} \label{eq:discretized_sys}
\Sigma_{\mathrm{LS}} \left\{
\begin{aligned}  
x_{k+1} &  = \hat{A}_{\sigma_k}x_{k}+\hat{B}_{\sigma_k}u_{k} \\
y_{k} & = Cx_{k}, \mbox{ } \forall k \in \mathbb{N}.
\end{aligned} 
\right.
\end{equation}
where $\sigma_k \in \mathbb{I}_1^{D}$ is called the value of the \emph{switching sequence} at time $k \in \mathbb{N}$. 

Models of the form \eqref{eq:discretized_sys} are a subclass of discrete-time LS SS representations where the read-out map represented by the matrix $C$ is constant and independent from the value of the switching signal $\sigma_k$ at each time instant $k \in \mathbb{N}$. Hence from now on, we will refer to the system representations of the form \eqref{eq:discretized_sys} as LS SS representations and formally define the tuple $\Sigma_{\mathrm{LS}}=(\{ (\hat{A}_i,\hat{B}_i,C) \}_{i=1}^{D})$ with $\hat{A}_i \in \mathbb{R}^{n \times n}$, $\hat{B}_i \in \mathbb{R}^{n \times m}$ for all $i \in \mathbb{I}_1^{D}$, $C \in \mathbb{R}^{p \times n}$ as an \emph{LS SS representation}. We remark that the discrete-time LS SS representation described by \eqref{eq:discretized_sys} \emph{completely} models the behavior of $\Sigma_{\mathrm{SD}}$ in sampling instants. More clearly, note that each linear mode $(\hat{A}_i,\hat{B}_i,C)$, $i \in \mathbb{I}_1^D$ corresponds to the $i$th element of the sampling interval set $\mathcal{H}=\{ \hat{h}_1, \dots, \hat{h}_D  \}$, i.e., if the $k$th sampling interval $h_k$, $k \in \mathbb{N}$ is chosen as $h_k=\hat{h}_i$, $i \in \mathbb{I}_1^D$, then the value of the switching signal at time instant $k$ is $\sigma_k=i$. In the following, analogous to the LTI case, $\dim(\Sigma_{\mathrm{LS}})$ will be used to denote the dimension $n$ of the state-space of $\Sigma_{\mathrm{LS}}$ and the number $n$ will be called the \emph{order} of $\Sigma_{\mathrm{LS}}$.

Now we can state the problem considered in the paper as follows. 

\textbf{Problem} Let $\Sigma_{\mathrm{LTI}}$ be a continuous-time LTI plant model of order $n$, which is to be sampled aperiodically with respect to the set $\mathcal{H}=\{ \hat{h}_1, \dots, \hat{h}_D \}$ to form the sampled data system $\Sigma_{\mathrm{SD}}$. Compute a discrete-time model $\bar{\Sigma}_{\mathrm{LS}}$ of order $r<n$ which is an approximation of the input-output behavior of $\Sigma_{\mathrm{SD}}$ in sampling instants. 

\begin{figure}[h!]
	\begin{center}		
		\begin{tikzpicture}
		\node at (-3.25,3.5) {\textbf{Approach 1}};
		\draw (-4,3) rectangle (-2.5,2);
		\draw (-4,1) rectangle (-2.5,0);
		\draw (-4,-1) rectangle (-2.5,-2);
		\draw (-4,-3) rectangle (-2.5,-4);
		\draw [->] (-3.25,2) -- (-3.25,1);
		\node at (-3,1.5) {\textbf{1}};
		\node at (-4.5,1.5) {\small{Model Reduction}};
		\draw [->] (-3.25,0) -- (-3.25,-1);
		\node at (-3,-0.5) {\textbf{2}};
		\node at (-4,-0.5) {\small{Sampling}};
		\draw [->] (-3.25,-2) -- (-3.25,-3);
		\node at (-3,-2.5) {\textbf{3}};
		\node at (-4.2,-2.5) {\small{LS Modeling}};
		\node at (-0.25,3.5) {\textbf{Approach 2}};
		\draw (-1,3) rectangle (0.5,2);
		\draw (-1,1) rectangle (0.5,0);
		\draw (-1,-1) rectangle (0.5,-2);
		\draw (-1,-3) rectangle (0.5,-4);
		\draw [->] (-0.25,2) -- (-0.25,1);
		\node at (0.5,1.5) {\small{Sampling}};
		\node at (-0.5,1.5) {\textbf{5}};
		\draw [->] (-0.25,0) -- (-0.25,-1);
		\node at (0.8,-0.5) {\small{LS Modeling}};
		\node at (-0.5,-0.5) {\textbf{6}};
		\draw [->] (-0.25,-2) -- (-0.25,-3);
		\node at (1,-2.5) {\small{Model Reduction}};
		\node at (-0.5,-2.5) {\textbf{7}};
		\node at (-3.25,2.5) {$\Sigma_{\mathrm{LTI}}$};
		\node at (-3.25,0.5) {$\bar{\Sigma}_{\mathrm{LTI}}$};
		\node at (-3.25,-1.5) {$\bar{\Sigma}_{\mathrm{SD}}$};
		\node at (-3.25,-3.5) {$\bar{\Sigma}_{\mathrm{LS}}$};
		\node at (-0.25,2.5) {$\Sigma_{\mathrm{LTI}}$};
		\node at (-0.25,0.5) {$\Sigma_{\mathrm{SD}}$};
		\node at (-0.25,-1.5) {$\Sigma_{\mathrm{LS}}$};
		\node at (-0.25,-3.5) {$\bar{\Sigma}_{\mathrm{LS}}$};
		\draw [->] (-2.5,0) .. controls (-1.75,-1) and (-1.75,-2) .. (-2.5,-3);
		\node at (-1.75,-1.5) {\textbf{4}};
		\draw [->] (-1,2) .. controls (-1.75,1) and (-1.75,0) .. (-1,-1);
		\node at (-1.75,0.5) {\textbf{8}};
		\end{tikzpicture}
		\caption{Overview of the two model reduction approaches.}
		\label{fig:overview_approaches}
	\end{center}
\end{figure}
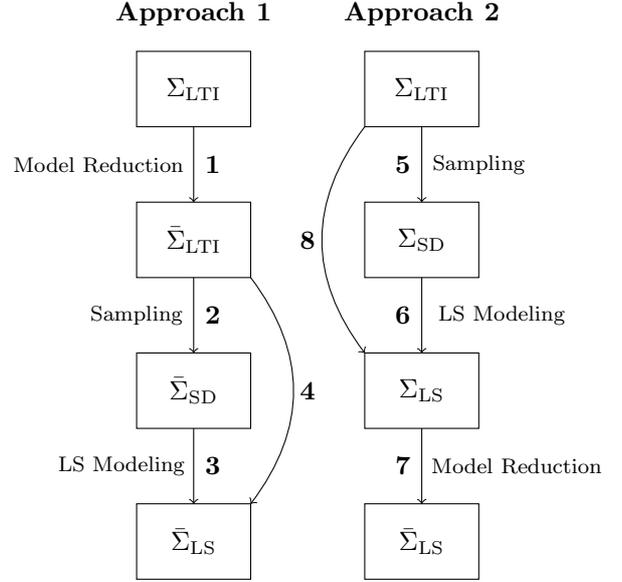

Two intuitive approaches (see Figure \ref{fig:overview_approaches}) can be proposed for the solution of this problem:

\textbf{Approach 1} Let $\Sigma_{\mathrm{LTI}}=(A,B,C)$ be the continuous-time LTI plant which is to be sampled aperiodically with respect to $\mathcal{H}=\{ \hat{h}_1, \dots, \hat{h}_D \}$. Compute from $\Sigma_{\mathrm{LTI}}$ another LTI SS model $\bar{\Sigma}_{\mathrm{LTI}}=(\bar{A},\bar{B},\bar{C})$ of order $r<n$ who approximates the input-output behavior of $\Sigma_{\mathrm{LTI}}$. Let $\bar{\Sigma}_{\mathrm{SD}}=(\bar{A},\bar{B},\bar{C},\mathcal{H})$ be the sampled data system corresponding to $\bar{\Sigma}_{\mathrm{LTI}}$. Compute from $\bar{\Sigma}_{\mathrm{SD}}$ the LS SS model $\bar{\Sigma}_{\mathrm{LS}}=(\{ (\hat{\bar{A}}_i,\hat{\bar{B}}_i,\bar{C}) \}_{i=1}^{D})$ of order $r<n$ of the form \eqref{eq:discretized_sys}.

\textbf{Approach 2} Let $\Sigma_{\mathrm{SD}}=(A,B,C,\mathcal{H})$ be the sampled data system with $\mathcal{H}=\{ \hat{h}_1, \dots, \hat{h}_D \}$ of the form \eqref{eq:sampled_plant} corresponding to the continuous-time LTI plant $\Sigma_{\mathrm{LTI}}=(A,B,C)$ of the form \eqref{eq:LTI_cont_sys}. Let $\Sigma_{\mathrm{LS}}=(\{ (\hat{A}_i,\hat{B}_i,C) \}_{i=1}^{D})$ of the form \eqref{eq:discretized_sys} be the corresponding LS SS model for the sampled data system $\Sigma_{\mathrm{SD}}$. Compute from  $\Sigma_{\mathrm{LS}}$ another LS SS model $\bar{\Sigma}_{\mathrm{LS}}=(\{ (\bar{\hat{A}}_i,\bar{\hat{B}}_i,\hat{C}) \}_{i=1}^{D})$ of order $r<n$ who approximates the input-output behavior of $\Sigma_{\mathrm{LS}}$.

\begin{Remark}
	Note that the symbol $\bar{}$ over a system representation $\Sigma$ is used for indicating that $\bar{\Sigma}$ is an approximation system for $\Sigma$; whereas the subscripts $\mathrm{LTI}$, $\mathrm{SD}$, $\mathrm{LS}$ are used for referring to the particular class of system representations of the form \eqref{eq:LTI_cont_sys}, \eqref{eq:sampled_plant} and \eqref{eq:discretized_sys} respectively. The symbol $\bar{}$ when used above a system matrix $A,B$ or $C$ of a system $\Sigma$, indicates that $\bar{A}$, $\bar{B}$ or $\bar{C}$ are the system parameters of $\bar{\Sigma}$. Finally, the symbol $\hat{}$ over a system parameter $A$ or $B$ of $\Sigma_{\mathrm{SD}}$ is used to indicate that $\hat{A}_i$ and $\hat{B}_i$ for all $i \in \mathbb{I}_1^D$ are the matrices of the form \eqref{eq:LSS_parameters} of the resulting LS SS representation corresponding to the sampled data system with $\mathcal{H}=\{ \hat{h}_1, \dots, \hat{h}_D \}$. The symbols $\hat{\bar{}}$ and $\bar{\hat{}}$ used above a system matrix in the definitions of \textbf{Approach 1} and \textbf{Approach 2} respectively can be interpreted with respect to this remark. 
\end{Remark}

\section{First Approach of Model Reduction} \label{sect:first_approach}

In this section, firstly we recall the concepts of Markov parameters and moment matching for LTI SS representations. Then we present \textbf{Approach 1} in detail.

\subsection{Review: Moment Matching for LTI SS Representations} \label{subsect:moment_match_LTI}

\begin{Notation}
	Let $a \in \mathbb{N} \backslash \{ 0 \}$. The set of \emph{continuous} and \emph{absolutely continuous} maps of the form $\mathbb{R}_+ \rightarrow \mathbb R^a$ is denoted by $\textbf{C}(\mathbb{R}_+,\mathbb R^a)$ and $\textbf{AC}(\mathbb{R}_+,\mathbb R^a)$ respectively; and the set of \emph{Lebesgue measurable maps} of the form $\mathbb{R}_+ \rightarrow \mathbb R^a$ which are integrable on any compact interval is denoted by $\textbf{L}_{\textbf{loc}}(\mathbb{R}_+, \mathbb R^a)$. 
\end{Notation}

We define the \emph{input-to-state map} $X^{x_0}_{\Sigma_{\mathrm{LTI}}}$ and \emph{input-to-output map} $Y^{x_0}_{\Sigma_{\mathrm{LTI}}}$ of a system $\Sigma_{\mathrm{LTI}}=(A,B,C)$ of the form \eqref{eq:LTI_cont_sys} as the maps 
\begin{align*}
X^{x_0}_{\Sigma_{\mathrm{LTI}}}: \textbf{L}_{\textbf{loc}}(\mathbb{R}_+,\mathbb{R}^m) & \rightarrow \textbf{AC}(\mathbb{R}_+,\mathbb{R}^n); \mbox{ } u \mapsto X^{x_0}_{\Sigma_{\mathrm{LTI}}}(u), \\ 
Y^{x_0}_{\Sigma_{\mathrm{LTI}}}: \textbf{L}_{\textbf{loc}}(\mathbb{R}_+,\mathbb{R}^m) & \rightarrow \textbf{C}(\mathbb{R}_+,\mathbb{R}^p); \mbox{ } u \mapsto Y^{x_0}_{\Sigma_{\mathrm{LTI}}}(u),
\end{align*}
defined by letting $t \mapsto X^{x_0}_{\Sigma_{\mathrm{LTI}}}(u)(t)$ be the solution to the first equation of \eqref{eq:LTI_cont_sys} with $x(0)=x_0$, and letting $Y^{x_0}_{\Sigma_{\mathrm{LTI}}}(u)(t)=C X^{x_0}_{\Sigma_{\mathrm{LTI}}}(u)(t)$ for all $t \in \mathbb{R}_+$ as in second equation of \eqref{eq:LTI_cont_sys}.

A moment's reflection lets us see that the $k$th Taylor series coefficient $M_{k}$ of $Y^{0}_{\Sigma_{\mathrm{LTI}}}$ around $t=0$ for the unit impulse input will be
\begin{equation} \label{rev1:markov}
M_{k}=CA^kB, \mbox{ } k \in \mathbb{N}
\end{equation}
where $A^0$ defined to be $I_n$. The coefficients $M_k$, $k \in \mathbb{N}$ are called the \emph{Markov parameters} or the \emph{moments} of the system $\Sigma_{\mathrm{LTI}}$. Hence, it is possible to approximate the input-output behavior of $\Sigma_{\mathrm{LTI}}$ by another system $\bar{\Sigma}_{\mathrm{LTI}}$ (possibly of reduced order), whose first some number of Markov parameters are equal to the corresponding ones of $\Sigma_{\mathrm{LTI}}$. If this number is chosen to be $N \in \mathbb{N}$, we will call such approximations as $N$-\emph{partial realizations} of $\Sigma_{\mathrm{LTI}}$. More precisely, a continuous-time LTI SS representation $\bar{\Sigma}_{\mathrm{LTI}}=(\bar{A},\bar{B},\bar{C})$ is an $N$-partial realization of another continuous-time LTI SS representation $\Sigma_{\mathrm{LTI}}=(A,B,C)$  if 
\begin{equation*}
	CA^kB = \bar{C}\bar{A}^k\bar{B} \mbox{, } k=0,\ldots,N. 
\end{equation*}

The problem of model reduction of LTI systems by moment matching can now be stated as follows. Consider an LTI system $\Sigma_{\mathrm{LTI}}=(A,B,C)$ of the form \eqref{eq:LTI_cont_sys} and fix $N \in \mathbb{N}$. Find another LTI system $\bar{\Sigma}_{\mathrm{LTI}}$ of order $r$ strictly less than $n$ such that $\bar{\Sigma}_{\mathrm{LTI}}$ is an $N$-partial realization of $\Sigma_{\mathrm{LTI}}$.

Below we recall a basic theorem on how to compute an $N$-partial realization for the LTI case. For this purpose, we define the $N$-partial reachability space of a continuous-time LTI realization $\Sigma_{\mathrm{LTI}}=(A,B,C)$ as
\begin{equation} \label{eq:N_reach_LTI}
\mathscr{R}^N_{\mathrm{LTI}} = \IM \left( \begin{bmatrix} B & AB & \cdots & A^{N}B \end{bmatrix} \right).
\end{equation}
for all $N \in \mathbb{N}$ with $A^0:=I_n$.

\begin{Theorem} \label{theo:mod_red_1}
	(Moment Matching for LTI SS Representations, [\cite{antoulas}]).
	Let $\Sigma_{\mathrm{LTI}}=(A,B,C)$ be a continuous-time LTI SS representation of the form \eqref{eq:LTI_cont_sys}, $N \in \mathbb{N}$ and $V \in \mathbb{R}^{n \times r}$ be a full column rank matrix such that
	\begin{equation*}
	\mathscr{R}^{N}_{\mathrm{LTI}} = \IM (V).
	\end{equation*}
	If $\bar{\Sigma}_{\mathrm{LTI}}=(\bar{A}, \bar{B}, \bar{C})$ is an LTI SS representation such that the matrices $\bar{A}$, $\bar{B}$, $\bar{C}$ are defined as
	\begin{equation} \label{eq:new_mat_parameters_LTI}
	\bar{A}=V^{-1}AV \mbox{, } \bar{B}=V^{-1}B \mbox{, } \bar{C}=CV,
	\end{equation}
	where $V^{-1}$ is a left inverse of $V$, then $\bar{\Sigma}_{\mathrm{LTI}}$ is an $N$-partial realization of $\Sigma_{\mathrm{LTI}}$.
\end{Theorem}

\subsection{Approach 1} \label{subsect:approach_1}

Now the first approach for model reduction of aperiodically sampled data systems can be stated in detail as follows:

\textbf{Approach 1} Let $\Sigma_{\mathrm{LTI}}=(A,B,C)$ be the continuous-time LTI plant of order $n$ which is to be sampled aperiodically with respect to $\mathcal{H}=\{ \hat{h}_1, \dots, \hat{h}_D \}$. By using Theorem \ref{theo:mod_red_1}, compute an $N$-partial realization  $\bar{\Sigma}_{\mathrm{LTI}}=(\bar{A},\bar{B},\bar{C})$ of $\Sigma_{\mathrm{LTI}}$ such that the order of $\bar{\Sigma}_{\mathrm{LTI}}$ is $r<n$. Let $\bar{\Sigma}_{\mathrm{SD}}=(\bar{A},\bar{B},\bar{C},\mathcal{H})$ be the sampled data system corresponding to $\bar{\Sigma}_{\mathrm{LTI}}$. Compute from $\bar{\Sigma}_{\mathrm{SD}}$ the LS SS model $\bar{\Sigma}_{\mathrm{LS}}=(\{ (\hat{\bar{A}}_i,\hat{\bar{B}}_i,\bar{C}) \}_{i=1}^{D})$ of order $r<n$ of the form \eqref{eq:discretized_sys}, with the procedure given in Section \ref{sect:sampled_to_LS}.

\begin{Corollary}[Theorem \ref{theo:mod_red_1}]
	The resulting reduced order LS SS model $\bar{\Sigma}_{\mathrm{LS}}$ computed by \textbf{Approach 1} corresponds to the discrete-time, time varying model $\Sigma_{\mathrm{disc}}$
	\begin{equation} \label{eq:approach1_red_ord_time_vary}
	\bar{\Sigma}_{\mathrm{disc}} \left\{
	\begin{aligned}
	\bar{x}_{k+1} & =\bar{\Phi} (h_k)\bar{x}_k + \bar{\Gamma} (h_k) u_k, \\
	\bar{y}_k & = \bar{C}\bar{x}_k, \mbox{ } \forall h_k \in \mathcal{H}, \mbox{ } \forall k \in \mathbb{N},
	\end{aligned} \right.
	\end{equation}
	of the sampled data system. In \eqref{eq:approach1_red_ord_time_vary}
	\begin{equation*}
	\bar{x}_k=V^{-1}x_k \mbox{, } \bar{\Phi}(h_k)=e^{\bar{A}h_k} \mbox{, } \bar{\Gamma}(h_k)=\left( \int_{0}^{h_k}e^{\bar{A}s}ds \right)\bar{B}.
	\end{equation*} 
\end{Corollary}

\section{Second Approach of Model Reduction} \label{sect:second_approach}

In this section, we recall the concepts of Markov parameters and moment matching for LS SS representations and the analogy between the LTI case. Then we present \textbf{Approach 2} in detail.

\subsection{Review: Moment Matching for LS SS Representations}

\begin{Notation}
	In the sequel, we use the following notation and terminology: If $s= s_0 \cdots s_N$ is a sequence with $N+1$ elements, $N \in \mathbb{N}$, we denote the number $N$ as $|s|=N$ and call $|s|$ as the length of the sequence $|s|$. We use $\textbf{Q}$ to denote the set of finite sequences in $Q=\{ 1, \dots, D \}$, $D \ge 1$, i.e., $\textbf{Q} = \{ \sigma=\sigma_0 \cdots \sigma_N \mid \sigma_0, \cdots, \sigma_N \in Q, N \in \mathbb{N} \}$; $\textbf{U}$ to denote the set of finite sequences in $\mathbb{R}^{m}$, i.e., $\textbf{U} = \{ u=u_0 \cdots u_N \mid u_0, \cdots, u_N \in \mathbb{R}^{m}, N \in \mathbb{N} \}$; $\textbf{X}$ to denote the set of finite sequences in $\mathbb{R}^{n}$, i.e., $\textbf{X} = \{ x=x_0 \cdots x_N \mid x_0, \cdots, x_N \in \mathbb{R}^{n}, N \in \mathbb{N} \}$ and $\textbf{Y}$ to denote the set of finite sequences in $\mathbb{R}^{p}$, i.e., $\textbf{Y} = \{ y=y_0 \cdots y_N \mid y_0, \cdots, y_N \in \mathbb{R}^{p}, N \in \mathbb{N} \}$. In addition, we will write $\overline{ \textbf{U} \times \textbf{Q} }=\{ (u,\sigma) \in \textbf{U} \times \textbf{Q} \mid |u|=|\sigma| \}$.
\end{Notation}

We define the \emph{input-to-state map} $X^{x_0}_{\Sigma_{\mathrm{LS}}}$ and \emph{input-to-output map} $Y^{x_0}_{\Sigma_{\mathrm{LS}}}$ of a system $\Sigma_{\mathrm{LS}}$ of the form \eqref{eq:discretized_sys} as the maps 
\begin{align*}
\overline{ \textbf{U} \times \textbf{Q} } \to \textbf{X};
(u,\sigma) \mapsto X^{x_0}_{\Sigma_{\mathrm{LS}}}(u,\sigma)=x,\\
\overline{ \textbf{U} \times \textbf{Q} } \to \textbf{Y};~ (u,\sigma)\mapsto Y^{x_0}_{\Sigma_{\mathrm{LS}}}(u,\sigma)=y,
\end{align*}
defined by letting $k \mapsto X^{x_0}_{\Sigma_{\mathrm{LS}}}(u,\sigma)_k$ be the solution to the first equation of \eqref{eq:discretized_sys} with $x(0)=x_0$, and letting $Y^{x_0}_{\Sigma_{\mathrm{LS}}}(u,\sigma)_k=C X^{x_0}_{\Sigma_{\mathrm{LS}}}(u,\sigma)_k$ for all $k \in \mathbb{N}$ as in second equation of \eqref{eq:discretized_sys}.

Using \eqref{eq:discretized_sys}, one can see that the coefficients appearing in the output of $\Sigma_{\mathrm{LS}}$ for any pair of input and switching sequences  $(u,\sigma) \in \overline{ \textbf{U} \times \textbf{Q} }$ are of the form
\begin{equation} \label{eq:Markov_0_LSS}
C\hat{B}_j, \mbox{ } j \in \mathbb{I}_{1}^{D}
\end{equation}
and
\begin{equation} \label{eq:Markov_M_LSS}
C \hat{A}_{k_1} \cdots \hat{A}_{k_M} \hat{B}_j; \mbox{ } k_1, \cdots, k_M, j \in \mathbb{I}_{1}^{D}, \mbox{ } M \in \mathbb{N} \backslash \{ 0 \}.
\end{equation}
Analogously to the linear case we will call the coefficients of the form \eqref{eq:Markov_0_LSS} and \eqref{eq:Markov_M_LSS} as the \emph{Markov parameters} of $\Sigma_{\mathrm{LS}}=(\{ (\hat{A}_i,\hat{B}_i,C) \}_{i=1}^{D})$. Specifically, we will call the Markov parameters of the form \eqref{eq:Markov_0_LSS} as the Markov parameters of length $0$ and the Markov parameters of the form \eqref{eq:Markov_M_LSS} as the Markov parameters of length $M$ for any $M \in \mathbb{N} \backslash \{ 0 \}$.

In [\cite{bastugACC2014}] and [\cite{bastug_IEEE_TAC_2015}] it is shown that similarly to the LTI case, it is possible to approximate the input-output behavior of $\Sigma_{\mathrm{LS}}$ by another LS SS representation $\bar{\Sigma}_{\mathrm{LS}}$ (possibly of reduced order), whose Markov parameters up to a certain length $N \in \mathbb{N}$ is equal with the corresponding ones of $\Sigma_{\mathrm{LS}}$ \footnote{Even though the results in [\cite{bastugACC2014}] and [\cite{bastug_IEEE_TAC_2015}] are stated in the continuous-time context, the analogous results on $N$-partial realizations of discrete-time LS SS representations are also valid. See [\cite{bastugCDC2015}] for an application of these results for the model reduction of affine LPV systems in the discrete-time context.}. Again, we will call such approximations as $N$-\emph{partial realizations} of $\Sigma_{\mathrm{LS}}$. More precisely, a discrete-time LS SS representation $\bar{\Sigma}_{\mathrm{LS}}=(\{ (\bar{\hat{A}}_i,\bar{\hat{B}}_i,\bar{C}) \}_{i=1}^{D})$ is an $N$-partial realization of another discrete-time LS SS representation $\Sigma_{\mathrm{LS}}=(\{ (\hat{A}_i,\hat{B}_i,C) \}_{i=1}^{D})$ if 
\begin{equation*}
C\hat{B}_j=\bar{C}\bar{\hat{B}}_j, \mbox{ } j \in \mathbb{I}_{1}^{D} 
\end{equation*}
and
\begin{equation*}
C \hat{A}_{k_1} \cdots \hat{A}_{k_M} \hat{B}_j = \bar{C} \bar{\hat{A}}_{k_1} \cdots \bar{\hat{A}}_{k_M} \bar{\hat{B}}_j
\end{equation*}
for all $k_1, \cdots, k_M, j \in \mathbb{I}_{1}^{D}$ and $M \in \mathbb{I}_{1}^{N}$. Note that an $N$-partial realization $\bar{\Sigma}_{\mathrm{LS}}$ of $\Sigma_{\mathrm{LS}}$ will have the \emph{same} output with $\Sigma_{\mathrm{LS}}$ for all time instants up to $N$, i.e., $k \in \mathbb{I}_0^N$, for all input and switching sequences. The reason why the output of an $N$-partial realization is indeed an approximation for the output of the original system model for also the time instants $k > N$ can be found in [\cite{bastugCDC2015}].

The problem of model reduction of LS systems by moment matching can now be stated as follows: Consider an LS SS model $\Sigma_{\mathrm{LS}}=(\{ (\hat{A}_i,\hat{B}_i,C) \}_{i=1}^{D})$ of the form \eqref{eq:discretized_sys} of order $n$ and fix $N \in \mathbb{N}$. Find another LS SS model $\bar{\Sigma}_{\mathrm{LS}}$ of order $r$ strictly less than $n$ such that $\bar{\Sigma}_{\mathrm{LS}}=(\{ (\bar{\hat{A}}_i,\bar{\hat{B}}_i,\bar{C}) \}_{i=1}^{D})$ is an $N$-partial realization of $\Sigma_{\mathrm{LS}}$.

Next, we recall a theorem of model reduction with $N$-partial realizations for the LS case [\cite{bastugACC2014}]. This theorem (Theorem \ref{theo:mod_red_2}) can be considered as the analogous of Theorem \ref{theo:mod_red_1} for the LS case (or in other words Theorem \ref{theo:mod_red_1} is a special case of \ref{theo:mod_red_2} when the LS system consists of only one LTI system). For this purpose, we define inductively the $N$-partial reachability space of a discrete-time LS SS representation $\Sigma_{\mathrm{LS}}=(\{ (\hat{A}_i,\hat{B}_i,C) \}_{i=1}^{D})$ as
\begin{equation} \label{eq:N_reach_LS}
\begin{aligned}
\mathscr{R}^0_{\mathrm{LS}} & = \SPAN \bigcup_{j \in \mathbb{I}_1^{D}} \IM(B_{j}), \\
\mathscr{R}^{N}_{\mathrm{LS}} & =\mathscr{R}^0_{\mathrm{LS}}+\sum_{k \in \mathbb{I}_{1}^{D}} \IM (A_k\mathscr{R}^{N-1}_{\mathrm{LS}}), \mbox{ } N \geq 1.
\end{aligned}
\end{equation}
for all $N \in \mathbb{N}$. In \eqref{eq:N_reach_LS} the summation operator must be interpreted as the Minkowski sum of vector spaces.

\begin{Theorem} \label{theo:mod_red_2}
	(Moment Matching for LS SS Representations, [\cite{bastugACC2014}]).
	Let $\Sigma_{\mathrm{LS}}=(\{ (\hat{A}_i,\hat{B}_i,C) \}_{i=1}^{D})$ be a discrete-time LS SS representation of the form \eqref{eq:discretized_sys}, $N \in \mathbb{N}$ and $V \in \mathbb{R}^{n \times r}$ be a full column rank matrix such that
	\begin{equation} \label{eq:mod_red_2_V_mat}
	\mathscr{R}^{N}_{\mathrm{LS}} = \IM (V).
	\end{equation}
	If $\bar{\Sigma}_{\mathrm{LS}}=(\{ (\bar{\hat{A}}_i,\bar{\hat{B}}_i,\bar{C}) \}_{i=1}^{D})$ is an LS SS representation such that for each $i \in \mathbb{I}_1^D$, the matrices $\bar{\hat{A}}_i,\bar{\hat{B}}_i,\bar{C}$ are defined as
	\begin{equation} \label{eq:new_mat_parameters_LS}
	\bar{\hat{A}}_i=V^{-1}\hat{A}_iV \mbox{, } \bar{\hat{B}}_i=V^{-1}\hat{B}_i \mbox{, } \bar{C}=CV,
	\end{equation}
	where $V^{-1}$ is a left inverse of $V$, then $\bar{\Sigma}_{\mathrm{LS}}$ is an $N$-partial realization of $\Sigma_{\mathrm{LS}}$.
\end{Theorem}

Note that the key of model reduction lies in the number of columns of the full column rank projection matrix $V \in \mathbb{R}^{n \times r}$ such that $r < n$. Choosing the number $N$ small enough such that the matrix $V$ satisfies the condition \eqref{eq:mod_red_2_V_mat} and it has $r < n$ columns, results in the reduced order $N$-partial realization $\bar{\Sigma}_{\mathrm{LS}}$ of order $r$. A simple algorithm with polynomial computational complexity to compute the matrix $V$ in Theorem \ref{theo:mod_red_2} is given in [\cite{bastugACC2014}]. Note also that the counterpart of Theorem \ref{theo:mod_red_2} can be given dually, using matrix representations of the $N$-unobservability space. These discussions are left out of this paper for simplicity and they can be found in detail in [\cite{bastugACC2014}].

\subsection{Approach 2} \label{subsect:approach_2}

Now the second approach for model reduction of aperiodically sampled data systems can be stated in detail as follows:

\textbf{Approach 2} Let $\Sigma_{\mathrm{SD}}=(A,B,C,\mathcal{H})$ be the sampled data system with $\mathcal{H}=\{ \hat{h}_1, \dots, \hat{h}_D \}$ of the form \eqref{eq:sampled_plant} corresponding to the continuous-time LTI plant $\Sigma_{\mathrm{LTI}}=(A,B,C)$ of the form \eqref{eq:LTI_cont_sys} of order $n$. Let $\Sigma_{\mathrm{LS}}=(\{ (\hat{A}_i,\hat{B}_i,C) \}_{i=1}^{D})$ of the form \eqref{eq:discretized_sys} be the corresponding LS SS model for the sampled data system $\Sigma_{\mathrm{SD}}$ computed with the procedure given in Section \ref{sect:sampled_to_LS}. By using Theorem \ref{theo:mod_red_2} compute an $N$-partial realization $\bar{\Sigma}_{\mathrm{LS}}=(\{ (\bar{\hat{A}}_i,\bar{\hat{B}}_i,\bar{C}) \}_{i=1}^{D})$ of order $r<n$ for $\Sigma_{\mathrm{LS}}$.

Since we have proposed computing an $N$-partial realization based on the model \eqref{eq:discretized_sys} of an aperiodically sampled system as a model reduction approach for it, it could be helpful to relate the definition of $N$-partial realization to the model \eqref{eq:time_varying_repr} of the sampled data system. The following corollary is the direct consequence of the definition of $N$-partial realizations in the switched case and the representations \eqref{eq:time_varying_repr} and \eqref{eq:discretized_sys}.

\begin{Corollary}[Theorem \ref{theo:mod_red_2}]
	The $N$-partial realization $\bar{\Sigma}_{\mathrm{LS}}$ of the original LS SS model $\Sigma_{\mathrm{LS}}$ of the sampled data system computed by \textbf{Approach 2} corresponds to the time-varying model
	\begin{equation} \label{eq:red_ord_time_varying}
	\bar{\Sigma}_{\mathrm{disc}} \left\{
	\begin{aligned}
	\bar{x}_{k+1} & =\bar{\Phi} (h_k)\bar{x}_k + \bar{\Gamma} (h_k) u_k, \\
	\bar{y}_k & = \bar{C}\bar{x}_k, \mbox{ } \forall h_k \in \mathcal{H}, \mbox{ } \forall k \in \mathbb{N},
	\end{aligned} \right.
	\end{equation}
	such that the outputs $y_k$ of $\Sigma_{\mathrm{disc}}$ in \eqref{eq:time_varying_repr} and $\bar{y}_k$ of $\bar{\Sigma}_{\mathrm{disc}}$ in \eqref{eq:red_ord_time_varying} for any input $u \in \textbf{U}$ will be the same for all $k \in \mathbb{I}_0^N$. In other words,
	\begin{equation*}
	C \Phi(h_k) \cdots \Phi(h_i) \Gamma(h_i) = \bar{C} \bar{\Phi}(h_k) \cdots \bar{\Phi}(h_i) \bar{\Gamma}(h_i),
	\end{equation*}
	for all $k \in \mathbb{I}_0^N$ and $i \in \mathbb{I}_0^k$, where $h_l \in \mathcal{H}$ for all $l \in \mathbb{I}_i^k$. The relationship between $\Sigma_{\mathrm{disc}}$ of \eqref{eq:time_varying_repr} and $\bar{\Sigma}_{\mathrm{disc}}$ of \eqref{eq:red_ord_time_varying} can be constructed by stating that
	\begin{equation*}
	\bar{x}_k=V^{-1}x_k \mbox{, } \bar{\Phi}(h_k)=V^{-1}\Phi(h_k)V \mbox{, } \bar{\Gamma}(h_k)=V^{-1}\Gamma(h_k)
	\end{equation*} 
	for all $h_k \in \mathcal{H}$ and $k \in \mathbb{N}$.
\end{Corollary}

\section{Conservation of Stability} \label{sect:conserv_stability}

In this section, we build the relationship between the stability of the original continuous-time LTI system $\Sigma_{\mathrm{LTI}}$ and the quadratic stability of the reduced order discrete-time LS SS model $\bar{\Sigma}_{\mathrm{LS}}$ computed with both approaches. More specifically, we will show that as long as the original continuous-time LTI system $\Sigma_{\mathrm{LTI}}$ is stable, the reduced order discrete-time LS SS representation $\bar{\Sigma}_{\mathrm{LS}}$ modeling the sampled data system computed by \textbf{Approach 1} or \textbf{Approach 2} will be quadratically stable. As the final result of this section, we will extend this conservation of stability argument for the representations of the form \eqref{eq:time_varying_repr} of aperiodically sampled data systems. We will start with presenting two technical lemmas for the purpose of presenting this result.

\subsection{Technical Lemmas} \label{sect:stab_tech_lemmas}
In the sequel, we denote the fact that a matrix $G$ is positive definite (resp. positive semi-definite, negative definite, negative semi-definite) with $G > 0$ (resp. $G \ge 0$, $G < 0$, $G \le 0$).

\begin{Definition}[Quadratic stability] \label{def:quadratic}~~~~
	Let $\Sigma_{\mathrm{LS}}=(\{ (\hat{A}_i,\hat{B}_i,C) \}_{i=1}^{D})$ be an LS SS representation of the form \eqref{eq:discretized_sys}. The LS SS representation $\Sigma_{\mathrm{LS}}$ is quadratically stable if and only if there exists a symmetric positive definite $P \in \mathbb{R}^{n \times n}$ such that 
	\begin{equation} \label{eq:quad_stab_def}
	\hat{A}_i^{\mathrm{T}}P\hat{A}_i-P < 0, \quad \forall i \in \mathbb{I}_1^D.
	\end{equation}
\end{Definition}

\begin{Lemma} \label{theo:LTI_to_LS_preserve_stability}
	Let $\Sigma_{\mathrm{LTI}}=(A,B,C)$ be an LTI SS representation. For any $\mathcal{H}=\{ \hat{h}_1, \dots, \hat{h}_D \}$ with $D \in \mathbb{N} \backslash \{ 0 \}$, the LS SS model $\Sigma_{\mathrm{LS}}=(\{ (\hat{A}_i,\hat{B}_i,C) \}_{i=1}^{D})$ of the sampled data system $\Sigma_{\mathrm{SD}}=(A,B,C,\mathcal{H})$ is quadratically stable if $\Sigma_{\mathrm{LTI}}$ is stable.	
\end{Lemma}

\begin{pf}
	The stability of $\Sigma_{\mathrm{LTI}}=(A,B,C)$ implies the stability of the autonomous system $\Sigma^{\mathrm{aut}}_{\mathrm{LTI}}=(A,0,0)$. Then there exists a $P > 0$, $P^{\mathrm{T}}=P$ and $V(x(t))=x(t)^{\mathrm{T}}Px(t)$ such that
	\begin{equation*}
		V(x(t)) < V(x(0)), \mbox{ } \forall x(0) \in \mathbb{R}^n, x(0) \ne 0, \forall t \in \mathbb{R}_+ \backslash \{ 0 \}.
	\end{equation*}
	Then for all $x(0) \in \mathbb{R}^n, x(0) \ne 0$ and for all $t \in \mathbb{R}_+ \backslash \{ 0 \}$
	\begin{equation} \label{eq:proof_stab_LTI_to_LS_1}
		x(t)^{\mathrm{T}}Px(t) - x(0)^{\mathrm{T}}Px(0) < 0.
	\end{equation}
	Replacing $x(t)$ with $e^{At}x(0)$ in \eqref{eq:proof_stab_LTI_to_LS_1} yields that for all $x(0) \in \mathbb{R}^n, x(0) \ne 0$ and for all $t \in \mathbb{R}_+ \backslash \{ 0 \}$
	\begin{equation} \label{eq:proof_stab_LTI_to_LS_2}
		x(0)^{\mathrm{T}} \left( e^{A^{\mathrm{T}}t} P e^{At} - P \right)x(0) < 0.
	\end{equation}
	In turn \eqref{eq:proof_stab_LTI_to_LS_2} implies
	\begin{equation*}
		 e^{A^{\mathrm{T}}\hat{h}_i} P e^{A\hat{h}_i} - P < 0, \mbox{ } \forall \hat{h}_i \in \mathcal{H}.
	\end{equation*}
	Using \eqref{eq:time_varying_repr} we conclude that
	\begin{equation*}
		\Phi^{\mathrm{T}}(\hat{h}_i) P \Phi(\hat{h}_i) - P < 0, \mbox{ } \forall \hat{h}_i \in \mathcal{H}.
	\end{equation*}
	The proof of the statement follows by noticing that $\Phi(\hat{h}_i)=\hat{A}_i$, for all $i \in \mathbb{I}_1^D$. 
\end{pf}

Lemma \ref{theo:LTI_to_LS_preserve_stability} establishes the connection between the stability of $\Sigma_{\mathrm{LTI}}$ and the quadratic stability of $\Sigma_{\mathrm{LS}}$. Hence the proof of conservation of stability in the directions of \textbf{4} and \textbf{8} on Figure \ref{fig:overview_approaches} is done. The following lemma establishes the remaining part of the conservation of stability argument stated in the beginning of this section (namely, in the directions of \textbf{1} and \textbf{7} on Figure \ref{fig:overview_approaches}). 

\begin{Lemma} \label{theo:column_preserve_stability}
	Let $\Sigma_{\mathrm{LS}}=(\{ (\hat{A}_i,\hat{B}_i,C) \}_{i=1}^{D})$ be a quadratically stable LS SS representation of the form \eqref{eq:discretized_sys} and $P > 0$ be a solution of \eqref{eq:quad_stab_def}. If the left inverse $V^{-1}$ of the matrix $V \in \mathbb{R}^{n \times r}$ in Theorem \ref{theo:mod_red_2}  is chosen as $V^{-1}=(V^\mathrm{T}PV)^{-1}V^\mathrm{T}P$, then $\bar{\Sigma}_{\mathrm{LS}}=(\{ (\bar{\hat{A}}_i,\bar{\hat{B}}_i,\bar{C}) \}_{i=1}^{D})$ in Theorem \ref{theo:mod_red_2} is also quadratically stable.		
\end{Lemma}
\begin{pf}
	See Appendix \ref{sect:appendix}.
\end{pf}

\begin{Remark}
	Note that even though Lemma \ref{theo:column_preserve_stability} is presented in this paper as a step on proving the stability of the reduced order models for the sampled data systems computed with \textbf{Approach 1} or \textbf{Approach 2}, on the condition of stability of the original plant; it can also be considered as an independent stability result for model reduction of discrete-time LS SS representations.
\end{Remark}

With Lemma \ref{theo:column_preserve_stability} we have established the connection between the quadratic stability of $\Sigma_{\mathrm{LS}}$ and $\bar{\Sigma}_{\mathrm{LS}}$ in the direction of $7$ in Figure \ref{fig:overview_approaches}. Using this proof, proving the counterpart argument in the direction of $1$ in Figure \ref{fig:overview_approaches} is trivial \footnote{After this statement, one must also remark that numerical issues related to the particular implementation of  the moment matching algorithm can cause instability in the reduced order model even in the linear case and even when the original system is stable, [\cite{antoulas}].}. Therefore, the first statement in the beginning of the section has been proven. 

\subsection{Main Stability Result} \label{sect:stab_main_result}

With the following theorem, we can relate the conservation of stability argument to the discrete-time, time varying representations of the reduced order aperiodically sampled data systems of the form \eqref{eq:approach1_red_ord_time_vary} and \eqref{eq:red_ord_time_varying} respectively.

\begin{Theorem}
	Let $\Sigma_{\mathrm{LTI}}=(A,B,C)$ be stable. Then 
	
	\emph{(i)} The model $\bar{\Sigma}_{\mathrm{disc}}$ of the form \eqref{eq:approach1_red_ord_time_vary} corresponding to $\bar{\Sigma}_{\mathrm{LS}}$ computed with \textbf{Approach 1} is quadratically stable.
	
	\emph{(ii)} The model $\bar{\Sigma}_{\mathrm{disc}}$ of the form \eqref{eq:red_ord_time_varying} corresponding to $\bar{\Sigma}_{\mathrm{LS}}$ computed with \textbf{Approach 2} is quadratically stable.
\end{Theorem}
\begin{pf}
	The proof of part \emph{(i)} follows directly from the subsequent application of Lemma \ref{theo:column_preserve_stability} and \ref{theo:LTI_to_LS_preserve_stability} to $\Sigma_{\mathrm{LTI}}$ (note that to apply Lemma \ref{theo:column_preserve_stability} and \ref{theo:LTI_to_LS_preserve_stability} to $\Sigma_{\mathrm{LTI}}=(A,B,C)$ in this order, one simply considers $\Sigma_{\mathrm{LTI}}$ as an LS SS representation consisting only of one LTI system $(A,B,C)$).
	
	The proof of part \emph{(ii)} follows directly from the subsequent application of Lemma \ref{theo:LTI_to_LS_preserve_stability} and \ref{theo:column_preserve_stability} to $\Sigma_{\mathrm{LTI}}$.
\end{pf}

\section{Numerical Examples} \label{sect:numerical_examples}

In this section, two generic numerical examples are presented to illustrate and compare the two proposed model reduction procedures \footnote{The implementation of the two approaches in \textsc{Matlab} is freely available (with the examples) for experimentation from https://sites.google.com/site/mertbastugpersonal/. The original system parameters used and reduced order system parameters computed can be obtained from the same site.}. 

\textbf{Example 1}

\begin{figure} 	
	\centering
	\includegraphics[width=0.5\textwidth]{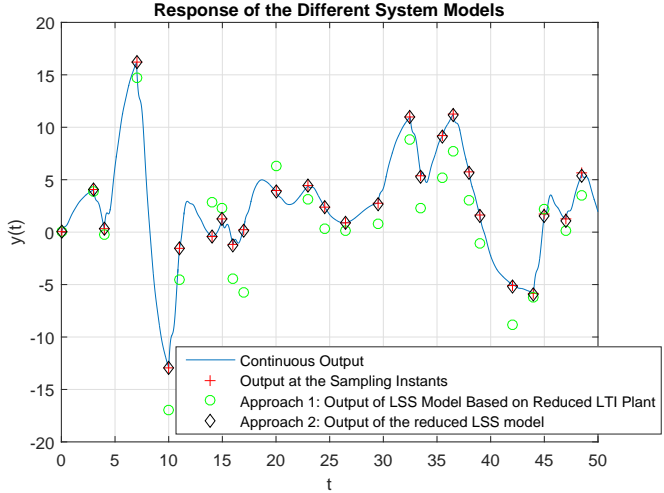}
	\caption{\textbf{Example 1}: The outputs resulting from the two approaches compared with the original output. For this simulation, the $\mathrm{BFR}$ of \textbf{Approach 1} is $53.8214\%$ and the one of \textbf{Approach 2} is $98.3700\%$.}
	\label{fig:example1}
\end{figure}

In the first example, the two approaches are applied to get a reduced order model for a single-input single-output (SISO), stable system $\Sigma_{\mathrm{LTI}}=(A,B,C)$ of order $50$, sampled to form the sampled data system $\Sigma_{\mathrm{SD}}=(A,B,C,\mathcal{H})$ with $\mathcal{H}=\{ \hat{h}_1, \hat{h}_2, \hat{h}_3, \hat{h}_4 \}=\{ 1, 1.5, 2, 3 \}$. For \textbf{Approach 1}, firstly a reduced order continuous-time LTI $17$-partial realization $\bar{\Sigma}_{\mathrm{LTI}}$ of $\Sigma_{\mathrm{LTI}}$ with order $18$ is computed. Then this model is used to get the reduced order (of order $18$) LS SS model $\bar{\Sigma}_{\mathrm{LS}}$ of the sampled data system $\Sigma_{\mathrm{SD}}$. For simulation, the output sequence $y_k$ of the original sampled data system $\Sigma_{\mathrm{SD}}$ and $\bar{y}_k$ of the reduced order LS model $\bar{\Sigma}_{\mathrm{LS}}$ are acquired for $k=\mathbb{I}_0^K$ where $K+1$ is the number of sampling instants of the simulation; by applying the same white Gaussian noise input sequence $u=u_0 \cdots u_K$, $u_k \in \mathcal{N}(0,1)$ for all $k \in \mathbb{I}_0^K$ and sampling sequence $h=h_0 \cdots h_K$, $h_k \in \mathcal{H}$ for all $k \in \mathbb{I}_0^K$. For this example, the total time horizon is chosen as $[0,50]$. For each simulation, the distance of the values $\bar{y}_k$ to the values $y_k$, $k \in \mathbb{I}_0^K$ are compared with the best fit rate ($\mathrm{BFR}$) [\cite{ljung}] which is defined as 
\begin{equation}
	\mathrm{BFR}=100 \% \max \left( 1-\frac{\sqrt{\sum_{k=0}^{K}\norm{y_k-\bar{y}_k}^2_2}}{\sqrt{\sum_{k=0}^{K} \norm{y_k-y_m}^2_2}},0 \right)
\end{equation}
where $y_m$ is the mean of the sequence $\{y_k\}_{k=0}^{K}$. The mean of the BFRs for $200$ such different simulations is acquired as $53.8303\%$ whereas the best BFR is acquired as $73.9027\%$ and the worst as $23.0697\%$. The output sequence $\{\bar{y}_k\}_{k=0}^{K}$ of the simulation giving the closest value to the mean of the BFRs over this $200$ simulations is illustrated in Figure \ref{fig:example1} together with the original output sequence $\{y_k\}_{k=0}^{K}$.

Then \textbf{Approach 2} is applied to the same example. Firstly the original LTI SS representation $\Sigma_{\mathrm{LTI}}$ is used to construct to LS SS representation $\Sigma_{\mathrm{LS}}$ which models the behavior of the sampled data system with respect to the sampling interval set $\mathcal{H}$. The model $\Sigma_{\mathrm{LS}}$ is then used to get the reduced order LS SS representation $\bar{\Sigma}_{\mathrm{LS}}$ using Theorem \ref{theo:mod_red_2}. The reduced order LS SS representation $\bar{\Sigma}_{\mathrm{LS}}$ in this case is a $2$-partial realization of order $18$. When the same $200$ simulations is done with this model with the specifications given for \textbf{Approach 1} of this example, the mean of the $\mathrm{BFR}$s over these simulations is $98.4222\%$, best $99.6449\%$, worst $95.5082\%$. Again, the output sequence giving the closest value to the mean of the BFRs over this $200$ simulations is illustrated in Figure \ref{fig:example1} together with the original output sequence.

\textbf{Example 2}

\begin{figure}
	\centering
	\includegraphics[width=0.5\textwidth]{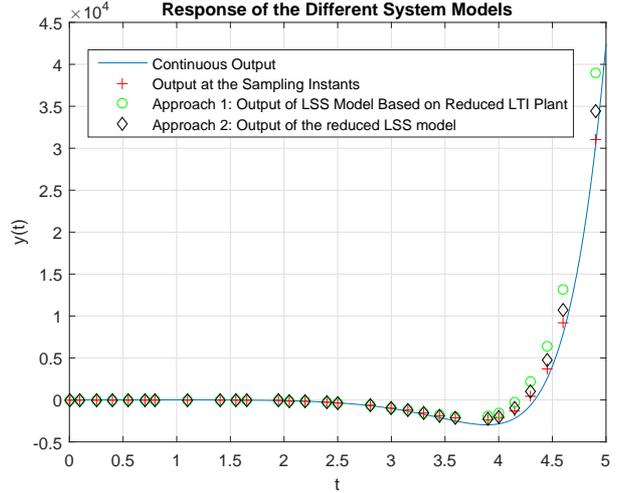}
	\caption{\textbf{Example 2}: The outputs resulting from the two approaches compared with the original output. For this simulation, the $\mathrm{BFR}$ of \textbf{Approach 1} is $91.4432\%$ and the one of \textbf{Approach 2} is $96.4294\%$.}
	\label{fig:example2}
\end{figure}

In the second example, the two approaches are applied to get a reduced order model for a SISO unstable system $\Sigma_{\mathrm{LTI}}=(A,B,C)$ of order $10$ sampled to form the sampled data system $\Sigma_{\mathrm{SD}}=(A,B,C,\mathcal{H})$ with $\mathcal{H}=\{ \hat{h}_1, \hat{h}_2, \hat{h}_3, \hat{h}_4 \}=\{ 0.1, 0.15, 0.2, 0.3 \}$. For \textbf{Approach 1}, a reduced order continuous-time LTI $3$-partial realization $\bar{\Sigma}_{\mathrm{LTI}}$ of $\Sigma_{\mathrm{LTI}}$ with order $4$ is computed. Then this model is used to get the reduced order (of order $4$) LS SS model $\bar{\Sigma}_{\mathrm{LS}}$ of the sampled data system $\Sigma_{\mathrm{SD}}$. The simulations are done with the input and sampling sequences with the specifications analagous to the ones given for \textbf{Example 1}. For this example, the total time horizon is chosen as $[0,5]$. The mean of the BFRs for $200$ simulations is acquired as $91.5237\%$ whereas the best BFR is acquired as $94.7476\%$ and the worst as $76.8753\%$. The output sequence $\{\bar{y}_k\}_{k=0}^{K}$ of the simulation giving the closest value to the mean of the BFRs over this $200$ simulations is illustrated in Figure \ref{fig:example2} together with the original output sequence.

Then \textbf{Approach 2} is applied to the same example. Firstly the original LTI SS representation $\Sigma_{\mathrm{LTI}}$ is used to construct to LS SS representation $\Sigma_{\mathrm{LS}}$ which models the behavior of the sampled data system with respect to the sampling interval set $\mathcal{H}$. The model $\Sigma_{\mathrm{LS}}$ is then used to get the reduced order LS SS representation $\bar{\Sigma}_{\mathrm{LS}}$ using Theorem \ref{theo:mod_red_2}. The reduced order LS SS representation $\bar{\Sigma}_{\mathrm{LS}}$ in this case is a $0$-partial realization of order $4$. When the same $200$ simulations is done with this model with the corresponding same input and sampling sequences used for \textbf{Approach 1}, the mean of the $\mathrm{BFR}$s is acquired $96.1276\%$, best $97.8198\%$, worst $91.2306\%$. Again, the output sequence giving the closest value to the mean of the BFRs over this $200$ simulations is illustrated in Figure \ref{fig:example2} together with the original output sequence.

\begin{table}[!t]
	\caption{The mean of $\mathrm{BFR}$s over $200$ simulations for both of the examples using \textbf{Approach 1} and \textbf{Approach 2}}
	\label{table1}
	\centering
	\begin{tabular}{|c|c|c|} 
		\hline
		 Ex. / App. & \textbf{Approach 1} & \textbf{Approach 2} \\ \hline
		\textbf{Example 1} & $53.8303\%$ & $98.4222\%$ 		 \\ \hline
		\textbf{Example 2} & $91.5237\%$ & $96.1276\%$ 	 \\
		\hline
	\end{tabular}
\end{table}

The results of the two simulations for both of the examples are summarized in Table \ref{table1} for \textbf{Approach 1} and \textbf{Approach 2}. Intuitively, the reason why \textbf{Approach 2} is superior to \textbf{Approach 1} for these examples can be explained as follows: \textbf{Approach 1} is based on model reduction of the original LTI plant where the sampling behavior is not considered at all. Whereas \textbf{Approach 2} applies the model reduction on the model which already takes into account the specific set of allowed sampling intervals. It should be noted that this statement may change depending on the specific example, since for the moment, no formal proof of comparison for the two methods can be given. 

\section{Conclusions}

Two approaches for model reduction of sampled data systems by moment matching is proposed. One approach relies on applying a classical model reduction by moment matching algorithm to the original LTI plant whereas the other relies on computing a reduced order model from the LS SS model of the sampled data system. With some numerical examples, the use of two approaches are illustrated and compared. For both approaches, it is shown that the stability of the original continuous-time LTI plant guarantees the quadratic stability of the resulting reduced order discrete-time LS SS model with respect to any finite allowed sampling interval set.


\bibliography{ifacconf}             
                                                   
\appendix

\section{Proof of Lemma \ref{theo:column_preserve_stability}} \label{sect:appendix}   
Below, we will use the following simple claims: 

\emph{(C1)} If $S \in \mathbb{R}^{n \times n}$ is symmetric negative (respectively positive) definite, then $\hat{S}=V^{\mathrm{T}}SV$ is also symmetric negative (respectively positive) definite.

\emph{(C2)} $V^{-1}=(V^\mathrm{T}PV)^{-1}V^\mathrm{T}P \implies V^{-1}P^{-1}(V^{-1})^{\mathrm{T}}=(V^\mathrm{T}PV)^{-1}$.

\emph{(CS)} (Schur Complement Lemma for positive/negative definiteness). Let $S \in \mathbb{R}^{n \times n}$ be a symmetric positive definite matrix and $G \in \mathbb{R}^{n \times n}$. Then $G^{\mathrm{T}}SG-S < 0 \iff GS^{-1}G^{\mathrm{T}}-S^{-1}< 0$.

Note that, by the assumption of the theorem, \eqref{eq:quad_stab_def} holds. Multiplying \eqref{eq:quad_stab_def} by $V^\mathrm{T}$ from left and $V$ from right for all $i \in \mathbb{I}_1^D$ and using \emph{(C1)} yields
\begin{equation*} 
V^\mathrm{T}\hat{A}_i^\mathrm{T}P\hat{A}_iV-V^\mathrm{T}PV<0 , \quad \forall i \in \mathbb{I}_1^D.
\end{equation*}
By \emph{(CS)} it follows that 
\begin{equation} \label{eq:proof_quad_stab_1}
\hat{A}_iV(V^\mathrm{T}PV)^{-1}V^\mathrm{T}\hat{A}_i^\mathrm{T} - P^{-1} <0 , \quad \forall i \in \mathbb{I}_1^D.
\end{equation}
In turn, multiplying \eqref{eq:proof_quad_stab_1} by $V^{-1}$ from left and $(V^{-1})^\mathrm{T}$ from right for all $i \in \mathbb{I}_1^D$ and using \emph{(C1)} yields
\begin{equation} \label{eq:proof_quad_stab_2}
V^{-1}\hat{A}_iV(V^\mathrm{T}PV)^{-1}V^\mathrm{T}\hat{A}_i^\mathrm{T}(V^{-1})^\mathrm{T} - V^{-1}P^{-1}(V^{-1})^\mathrm{T} <0,
\end{equation}
for all $i \in \mathbb{I}_1^D$. Using \emph{(C2)} and choosing $\bar{P}=V^\mathrm{T}PV$, the inequality \eqref{eq:proof_quad_stab_2} can be rewritten as
\begin{equation} \label{eq:proof_quad_stab_3}
\bar{\hat{A}}_i \bar{P}^{-1} \bar{\hat{A}}_i^\mathrm{T} - \bar{P}^{-1} <0 , \quad \forall i \in \mathbb{I}_1^D.
\end{equation}
Finally, using \emph{(CS)} one more time for \eqref{eq:proof_quad_stab_3} yields
\begin{equation} \label{eq:proof_quad_stab_4}
\bar{\hat{A}}_i^\mathrm{T} \bar{P} \bar{\hat{A}}_i - \bar{P} <0 , \quad \forall i \in \mathbb{I}_1^D.
\end{equation}
Since $\bar{P}=V^\mathrm{T}PV$ is symmetric and positive definite by \emph{(C1)}, \eqref{eq:proof_quad_stab_4} proves the quadratic stability of $\bar{\Sigma}_{\mathrm{LS}}=(\{ (\bar{\hat{A}}_i,\bar{\hat{B}}_i,\bar{C}) \}_{i=1}^{D})$.

\end{document}